\documentclass[aps,prd,a4paper,twocolumn,amsmath,showpacs,superscriptaddress,nofootinbib,preprintnumbers]{revtex4-1}

\usepackage{verbatim}
\usepackage[T1]{fontenc}
\usepackage[utf8]{inputenc}
\usepackage[american]{babel}
\usepackage{epsfig}
\usepackage{graphicx}
\usepackage{booktabs}
\usepackage{multirow}
\usepackage{dcolumn}
\usepackage{amsmath}
\usepackage{mathtools}
\usepackage{amsfonts}
\usepackage{amssymb}
\usepackage{epstopdf}
\usepackage{bm}
\usepackage{siunitx}
\usepackage{braket}
\usepackage{enumitem}
\usepackage{soul}
\usepackage[table]{xcolor}
\usepackage{color}
\usepackage{transparent}
\usepackage{pifont}

\definecolor{navyblue}{rgb}{0.0, 0.0, 0.5}
\definecolor{royalblue}{rgb}{0.25, 0.41, 0.88}
\definecolor{cadmiumgreen}{rgb}{0.0, 0.42, 0.24}
\definecolor{blue-violet}{rgb}{0.54, 0.17, 0.89}
\definecolor{darkviolet}{rgb}{0.58, 0.0, 0.83}
\definecolor{orange(colorwheel)}{rgb}{1.0, 0.5, 0.0}

\usepackage{hyperref}
\hypersetup{
    colorlinks=true, % false: boxed links; true: colored links
    linkcolor=royalblue, % color of internal links (change box color with linkbordercolor)
    citecolor=magenta}

\newcommand\ee{\end{equation}}
\newcommand\be{\begin{equation}}
\newcommand\eea{\end{eqnarray}}
\newcommand\bea{\begin{eqnarray}}

\newcommand{\neff}{N_{\rm eff}}

\newcommand{\om}{\Omega_m}

\newcommand{\lcdm}{\Lambda\mathrm{CDM}}

\newcommand{\alens}{A_{\rm lens}}
\newcommand{\alensz}{A_{{\rm lens},0}}

\usepackage{booktabs}
\usepackage{multirow}
\usepackage{dcolumn}
\usepackage{colortbl}

\definecolor{magenta(process)}{rgb}{1.0, 0.0, 0.56}

\definecolor{darkspringgreen}{rgb}{0.09, 0.45, 0.27}

\definecolor{royalblue(web)}{rgb}{0.25, 0.41, 0.88}
\begin{document}

\title{A Vacuum Phase Transition Solves $H_0$ Tension}  

\author{Eleonora Di Valentino}
\email{eleonora.divalentino@manchester.ac.uk}
\affiliation{Jodrell Bank Center for Astrophysics, School of Physics and Astronomy, University of Manchester, Oxford Road, Manchester, M13 9PL, UK}
\author{Eric V.\ Linder}
\email{evlinder@lbl.gov}
\affiliation{Berkeley Center for Cosmological Physics \& Berkeley Lab, University of California, Berkeley, CA 94720, USA}
\affiliation{Energetic Cosmos Laboratory, Nazarbayev University, Astana, Kazakhstan 010000} 
\author{Alessandro Melchiorri}
\email{alessandro.melchiorri@roma1.infn.it}
\affiliation{Physics Department and INFN, Universit\`a di Roma ``La Sapienza'', Ple Aldo Moro 2, 00185, Rome, Italy} 

\date{\today}

\preprint{}
\begin{abstract}
Taking the Planck cosmic microwave background data and the more direct Hubble constant measurement data as unaffected by systematic offsets, the values of the Hubble constant $H_0$ interpreted within the $\Lambda$CDM cosmological constant and cold dark matter cosmological model are in $\sim 3.3 \sigma$ tension. We show that the Parker vacuum metamorphosis model, physically motivated by quantum gravitational effects and with the same number of parameters as $\lcdm$, can remove the $H_0$ tension, and can give an improved fit to data (up to $\Delta\chi^2=-7.5$). It also ameliorates tensions with weak lensing data and the high redshift Lyman alpha forest data. We separately consider a scale dependent scaling of the gravitational lensing amplitude, such as provided by modified gravity, neutrino mass, or cold dark energy, motivated by the somewhat different cosmological parameter estimates for low and high CMB multipoles. We find that no such scale dependence is preferred. 
\end{abstract}

\maketitle

%%%%%%%%%%%%%%%%%%%%%
\section{Introduction}

Cosmic microwave background (CMB) measurements provide highly precise probes of the conditions and energy components of the universe over the entire age of the universe. Moreover, they can reveal the total age and scale of the universe, and so the present Hubble constant $H_0$. The Hubble constant can 
also be determined through local distance measurements, e.g.\ through crosscalibration of Cepheid and Type Ia supernova distances \cite{freedman,R16}. 
The latest values from these two methods, within the concordance $\lcdm$ model with a cosmological constant plus cold dark matter, are in $\sim 3\sigma$ tension. 
This is probably the most relevant tension present between current cosmological data sets and several works have recently appeared  discussing it or proposing different theoretical mechanisms as solution (see e.g.\ \cite{Zhao:2017urm,Yang:2017amu,Prilepina:2016rlq,Santos:2016sog,Kumar:2016zpg,joudaki,arman,Karwal:2016vyq,Ko:2016uft,Bernal:2016gxb,Qing-Guo:2016ykt,Zhang:2017idq,zhao,brust,ersola}).

Taking each set of cosmological data at face value (cf.~\cite{follinknox,feeney,zhangH0, Dhawan:2017ywl} regarding local $H_0$), we found in \cite{paper1} that the $H_0$ values could be consistent in a parameter space expanded to include further, not unreasonable, cosmological physics. 
In particular, altering the mechanism for cosmic acceleration from a 
cosmological constant to a particular form of dynamical dark energy would remove the tension. However, the form of dark energy required was quite unusual, not corresponding to the usual scalar field dark energy models. 
It needed to be phantom, with equation of state parameter $w<-1$, and moreover be rapidly evolving. 

These properties generally are not held simultaneously since they	
tend to	exacerbate problems of fine tuning and stability. However,
there is a model considered in the early days of dark energy investigations that possesses just these phenomenological properties, from a sound
theoretical foundation:	the vacuum metamorphosis (VM) model of	
\cite{parker,vm2,vm3}, which has a phase transition in the nature of the vacuum.	
In this	article	we explore the observational viability of VM in	fitting	the data simultaneously	and removing the tension in $H_0$ values. 

Another peculiarity in the data is that cosmological parameters estimated from small scales (CMB multipoles $\ell\gtrsim 1000$) are somewhat offset relative to the values estimated from large scales ($\ell\lesssim 1000$) \cite{Addison:2015wyg,1706.10286,elldivide2,elldivide3}. In particular, larger scales show some preference for a higher Hubble constant. We therefore separately explore cosmology fitting in a $\lcdm$ parameter space extended to allow for a scale dependent CMB lensing parameter $\alens$, reflecting some (unspecified) nonstandard scale dependent physics. 

Section~\ref{sec:vm} introduces	the VM model and lays out the foundation for using it with CMB and distance data. In Sec.~\ref{sec:fit} we present the cosmology fitting data and procedure. We carry out Markov chain Monte Carlo (MCMC) fits to the data for the VM model in Sec.~\ref{sec:vmfit} in the baseline and the extended parameter spaces, and discuss the results. 
In Sec.~\ref{sec:scale} we investigate an  alternative	approach to addressing the tension through	the use	of a scale dependent
$\alens$. We conclude in Sec.~\ref{sec:concl}.

%%%%%%%%%%%%%%%%%%%%%%%%%%%%%%%%%%%%%% 
\section{Vacuum Metamorphosis} \label{sec:vm} 

\subsection{Background} 

The two main data sets in tension on the value of $H_0$ are the CMB data from the Planck satellite \cite{Aghanim:2015xee} and the distance measurements from \cite{R16}, hereafter called R16. 
Taking the Planck+R16 constraints in the $w_0$--$w_a$ plane at face value, \cite{papero} found that they prefer the phantom region $w<-1$ and more deeply phantom in the past 
($w_a<0$, \cite{paper1}). A single canonical scalar field cannot achieve this, and even more complicated, and effectively arbitrary, fields have difficulty. 
While adding the JLA supernova constraints \cite{JLA} tends to shift the 
preferred area out of the phantom region, and adding baryon acoustic 
oscillation (BAO) data \cite{beutler2011,ross2014,anderson2014} tends to prefer less negative values of $w_a$, adding weak lensing or CMB lensing preserves the 
preference for deep phantom models. Here we mostly focus on just the 
Planck or Planck+R16 data sets. 

It is interesting to consider whether a reasonably physically motivated 
model can be found for this unusual region. The answer is yes: one of the earliest dark energy models, vacuum metamorphosis \citep{parker,vm2,vm3} 
lives in just this part of phase space. This model has a sound physical 
foundation, taking into account quantum loop corrections to	gravity in 
the presence of a massive scalar field. In the	
first order calculations, this gives rise to $R^2$ terms familiar
from, e.g., Starobinsky	gravity	and inflation \cite{staro}, where $R$ is  the Ricci scalar, but	Parker and collaborators were able to nonperturbatively sum the infinite series (under certain	restrictions) and find a closed	form solution.	

This solution indicates a phase transition in gravity similar to Sakharov's induced gravity \cite{sakharov}. 
The phase transition is induced once the Ricci scalar curvature $R$ has evolved to become of order the mass squared of the field, and thereafter $R$ is frozen to be of order $m^2$. 
This original model had one free parameter, $m^2$, which determined the 
matter density today, $\om$, giving it the same number of free parameters 
as flat $\lcdm$. 

Some later elaborations added a vacuum expectation value, somewhat inorganically, acting as a 
cosmological constant, but we will focus on the original, more 
elegant VM model. 

%%%%%%%%%%%%%%%%%%%%%%%%%%%%% 
\subsection{Relation to $w_0$--$w_a$}

A first question might be how to connect the observational motivation
for a particular region in the dark energy equation of state phase
space $w_0$--$w_a$, where $w_0$ is the present value of the equation
of state function $w(a)$ and $w_a$ a measure of its time variation,
to the theoretical VM model. It has been well established that the
$w_0$--$w_a$ parametrization provides an excellent fit (at the 0.1\%
level in observables) to a broad range of scalar field models
\cite{lin2003,calde}, but VM has a very rapid time evolution and is
not a standard scalar field model.

\begin{figure}[htbp!]
\includegraphics[width=\columnwidth]{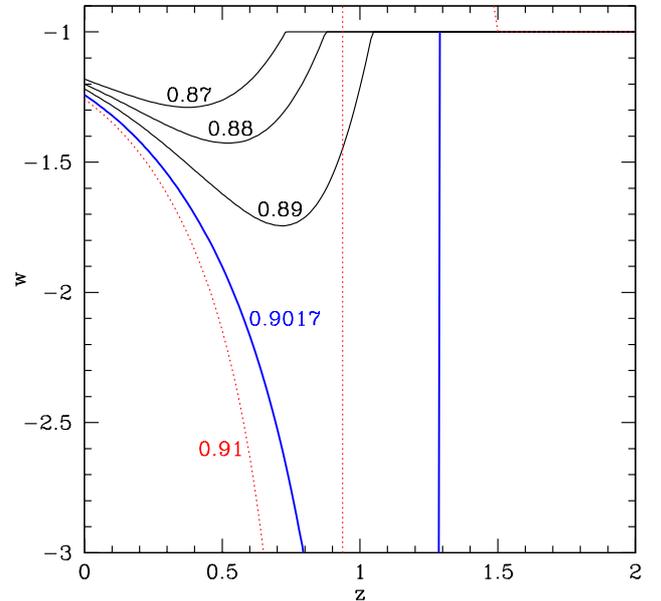}
\caption{
The effective dark energy equation of state evolution is plotted
vs redshift for several values of the mass parameter $M$, for $\om=0.3$. 
The bold blue
curve shows the original case (our preferred model) 
where there is no cosmological constant,
while the medium black curves show the elaborated case with an added
cosmological constant, and the dotted red curve shows one with a negative
cosmological constant (causing $w$ to first shoot up to large positive
values before it plummets to highly negative values).
}
\label{fig:vmwa}
\end{figure}

In Fig.~\ref{fig:vmwa} we illustrate the equation of state behavior
for the original, and some elaborated, VM models. One clearly sees
the phase transition at a fairly recent redshift, where the dark
energy deviates from an effective cosmological constant behavior of
$w=-1$ (for the elaborated cases) or newly appears in the phase
transition (in the original case). After the transition the dark energy
is highly phantom ($w<-1$) and then rapidly evolves toward $w=-1$
(with $w_a$ strongly negative) and an eventual de Sitter state as the
Ricci scalar freezes to the value of the field mass squared.

\begin{figure}[htbp!]
\includegraphics[width=\columnwidth]{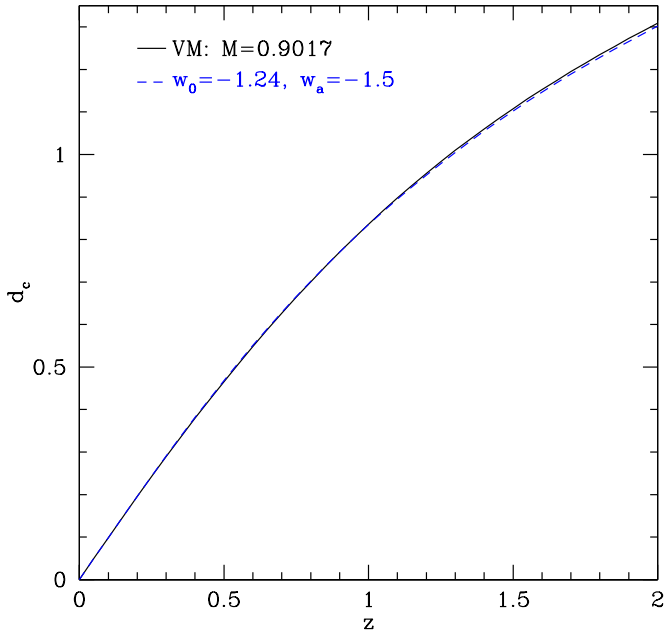}
\caption{The distance-redshift relation for the vacuum metamorphosis 
model without a cosmological constant -- the fastest evolving one -- 
is well fit by a standard $w_0$--$w_a$ model. Here the comoving distance, 
which enters the CMB distance to last scattering, and weak lensing, BAO, 
and supernova observations, is plotted vs redshift. 
}
\label{fig:vmdist}
\end{figure}

Even for the rapidly evolving case of no cosmological constant (our preferred case), 
the observational  implications of the model are well described by the standard $w_0$--$w_a$ 
parametrization since the phantom nature means that dark energy diminishes 
quickly into the past. Figure~\ref{fig:vmdist} illustrates the goodness 
of fit of the equivalent $w_0$--$w_a$ model for the most extreme case, 
that without a cosmological constant. The agreement in the distance-redshift 
relation is better than 
0.55\% at all redshifts (0.2\% in the distance to CMB last scattering), 
sufficient for current data precision. 
Note that $w_0=-1.24$, $w_a=-1.5$ is a good fit (lying near 68\% CL) to 
the Planck+R16 data, as well as when adding weak lensing or CMB lensing 
or shifting the local distance $H_0$ prior not lower than 70, as seen in 
\cite{paper1}.

%%%%%%%%%%%%%
\subsection{VM equations} 

The phase transition criticality condition is 
\be 
R=6(\dot H+H^2)=m^2 \ , 
\ee 
and, defining $M=m^2/(12H_0^2)$, the expansion behavior above and below 
the phase transition is 
\bea 
H^2/H_0^2&=&\Omega_m (1+z)^3+\Omega_r (1+z)^4\notag\\ 
&\quad&+M\left\{1-\left[3\left(\frac{4}{3\Omega_m}\right)^4 M(1-M)^3\right]^{-1}
\right\} \, z>z_t\\ 
H^2/H_0^2&=&(1-M)(1+z)^4+M\, , \quad z\le z_t \ . 
\eea 
The phase transition occurs at 
\be 
z_t=-1+\frac{3\Omega_m}{4(1-M)} \ , \label{eq:zt}
\ee 
(for simplicity of the expression we ignore the contribution of 
radiation energy density $\Omega_r$ at $z\lesssim1$). 

We see that above the phase transition, the universe behaves as one with 
matter plus a cosmological constant, and after the phase transition it 
effectively has a dark radiation component (the matter is hidden within 
this expression) that rapidly redshifts away leaving a 
de Sitter phase. The original model did not include an explicit high 
redshift cosmological constant; we see that this implies that 
\be 
\Omega_m=\frac{4}{3}\left[3M(1-M)^3\right]^{1/4} \ . \label{eq:vmom}
\ee 
So there is only one free parameter in the original model, either $M$ or $\Omega_m$, the same number as in $\lcdm$. 
For example, $\Omega_m=0.3$ implies $M=0.9017$. We emphasize that the de Sitter 
behavior at late times is not a result of a cosmological constant, but 
rather the intrinsic physics of the model. 

The effective dark energy equation of state (i.e.\ of the effective 
component once the matter contribution has been accounted for) is 
\be 
w(z)=-1-\frac{1}{3}\frac{3\Omega_m (1+z)^{3}-4(1-M)(1+z)^{4}}{M+(1-M)(1+z)^{4}-\Omega_m (1+z)^{3}} \ , 
\ee 
below the phase transition, and simply $w(z>z_t)=-1$ above the phase 
transition. In the case without a cosmological constant, there is no 
dark energy above the transition. 

The equation of state behavior is phantom, and more deeply phantom as 
the cosmological constant diminishes, as seen in Figure~\ref{fig:vmwa}. 
Note that for $M>0.9017$ (in the $\om=0.3$ case), the cosmological constant can go negative, and 
this leads initially to a highly positive equation of state just after 
the transition. This is not an observationally viable region. As $M$ falls below the 
critical value, the cosmological constant smooths out the rapid time 
variation, leading to a nearly constant $w(a)$. If $M$ falls too low, then the transition occurs in the future (see Eq.~\ref{eq:zt}), and we have simply the $\lcdm$ model for the entire history to the present. Moreover, $M$ then becomes no longer a free parameter but is given in terms of $\om$ by the requirement that $H(z=0)/H_0=1$. Thus, when considering the elaborated VM model with a free parameter $M$ we put a prior ranging between the lower and upper bounds, corresponding to $z_t\ge0$ and $\Omega_{\rm de}(z>z_t)\ge0$ respectively. But again, we regard 
the original VM model without cosmological constant as the most 
elegant and theoretically compelling.

%%%%%%%%%%%%%%%%%%%%%%%%%%%%%%
\section{Cosmological Parameter Fitting} \label{sec:fit} 

In order to study the vacuum metamorphosis model we consider a baseline parameter set plus extended scenarios. For our baseline, we consider $7$ cosmological parameters: the vacuum metamorphosis scale $M$, and the six parameters of the standard analysis, i.e.\ the baryon energy density $\Omega_bh^2$, the cold-dark-matter energy density $\Omega_ch^2$, the ratio between the sound
horizon and the angular diameter distance at decoupling $\theta_{s}$, the amplitude and spectral index of the primordial scalar perturbations $A_s$ and $n_s$ (at pivot scale $k_0=0.05 h\,{\rm Mpc}^{-1}$), and the reionization optical depth $\tau$. All these parameters are varied in a range of external, conservative, priors listed in Table~\ref{priors}. For the original VM model, 
$M$ is fixed by $\om$ (or v.v.) and so there are 6 parameters, as in $\lcdm$. 

We also consider two more extended scenarios in addition to our baseline model for testing VM. In the first scenario we add variations in $3$ more parameters: the total neutrino mass for the $3$ standard neutrinos $\Sigma m_\nu$, the running of the scalar spectral index $dn_s/d\ln k$, and the effective number of relativistic degrees of freedom $N_{\rm eff}$. Finally, in the last scenario, we also consider variation in the gravitational lensing
amplitude $\alens$ of the CMB angular power spectra (see e.g.\ \cite{calens}). This scales the CMB lensing strength on all scales by a constant, relative to the prediction of the model being considered. 

We analyze these cosmological parameters by making use of the high-$\ell$ temperature and low-$\ell$ temperature and polarization CMB angular power spectra released by Planck 2015 \cite{Aghanim:2015xee}. We refer to this dataset as ``Planck TT'', and it includes the large angular-scale temperature and polarization anisotropy measured by the Planck LFI experiment and the small-scale temperature anisotropies measured by Planck HFI.
Moreover, we add to Planck TT the high-$\ell$ polarization data measured by Planck HFI \cite{Aghanim:2015xee}, and we refer to this dataset simply as ``Planck''. This is our baseline data. 
We sometimes also consider the ``R16'' dataset in the form of an external Gaussian prior on the Hubble constant $H_0=73.24\pm1.74$ km/s/Mpc at $68 \%$ c.l., as measured by \cite{R16}.

In order to derive constraints on the parameters, we use the November 2016 version of the publicly available Monte Carlo Markov Chain package \texttt{cosmomc} \cite{Lewis:2002ah}. This code has a convergence diagnostic based on the Gelman and Rubin statistic and includes the support for the Planck data release 2015 Likelihood Code \cite{Aghanim:2015xee} (see \url{http://cosmologist.info/cosmomc/}), implementing an efficient sampling by using the fast/slow parameter decorrelations \cite{Lewis:2013hha}.
We also consider the impact of CMB foregrounds by including additional nuisance parameters and marginalizing over them as described in \cite{Aghanim:2015xee} and \cite{planck2015}.

\begin{table}
\begin{center}
\begin{tabular}{c|c}
% \hline
Parameter                    & Prior\\
\hline 
%\hspace{1mm}\\ 
$\Omega_{\rm b} h^2$         & $[0.005,0.1]$\\
$\Omega_{\rm c} h^2$       & $[0.001,0.99]$\\
$\tau$                       & $[0.01,0.8]$\\
$n_s$                        & $[0.8, 1.2]$\\
$\log[10^{10}A_{s}]$         & $[2,4]$\\
$\Theta_{\rm s}$             & $[0.5,10]$\\ 
%$M_{\rm vacuum}$ & [0,0.9]\\ 
$M$ & [$M_{\rm low}$,$M_{\rm high}$]\\
$\sum m_\nu$ (eV)               & $[0,5]$\\
$N_{\rm eff}$ & [0.05,10]\\ 
$\frac{dn_s}{d\ln k}$ & [-1,1]\\
$\alens$ & [0,10]\\
$B$ & [-0.4,0.4] \\
%$n$                        & $[-0.1, 0.1]$\\
% \hline
\end{tabular}
\end{center}
\caption{Flat priors on the various cosmological parameters used in this paper. $M_{\rm low}$ and $M_{\rm high}$ 
are given by conditions described in the text on Eqs.~(\ref{eq:zt}) and (\ref{eq:vmom}) respectively, as functions of $\om$. 
}
\label{priors}
\end{table}

%%%%%%%%%%%%%%%%%%%%%%%%%%%%%%%%%%%
\section{Vacuum Metamorphosis Cosmology Fits} \label{sec:vmfit} 

\subsection{Original VM}

%%%%%%%%%%%%%%%%%%%%%%%%%%%%%
\begin{figure*}
\centering
\includegraphics[width=2\columnwidth]{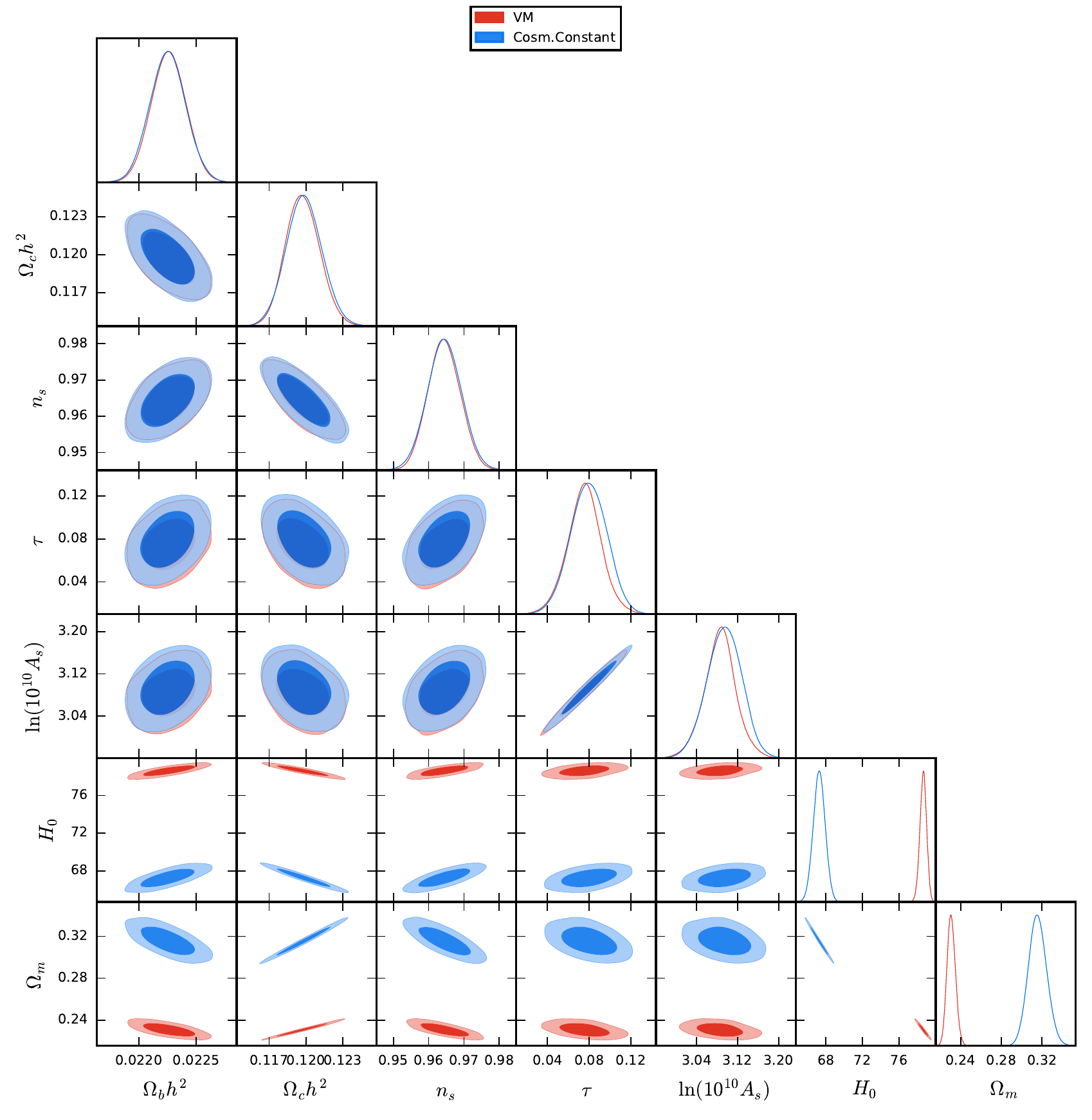} 
\caption{Triangular plot showing the posteriors of the cosmological parameters for $\lcdm$ and the original VM model, along with their 2D joint confidence contour at 68\% CL and 95\% CL. This is for baseline CMB data only, in the $6$ parameter space.}
\label{figtriM}
\end{figure*}
%%%%%%%%%%%%%%%%%%%%%%%%%

To begin, we consider the original VM model without cosmological constant. This has the same number of dark energy parameters as the standard  $\lcdm$ case, and as we know from Sec.~\ref{sec:vm}, 
it is also consistent with the region in the $w_0$--$w_a$ phase space 
preferred by the CMB data. The constraints on cosmological parameters in the case of variation of the standard $6$ parameters are reported in Table~\ref{tablefix} for different choices of datasets.
As we can see, assuming VM can indeed raise the Hubble constant but in fact it overshoots the R16 value, with Planck data alone providing a constraint $H_0=78.61\pm0.38$ (see Table~\ref{tablefix}). This, in practice, replaces one $3\sigma$ tension with its opposite.

The VM model and $\lcdm$ give similar results for most of the parameters, except for $H_0$ and $\om$ (and $\sigma_8$ which depends on $\om$).  This is clearly exhibited in Fig.~\ref{figtriM} where we report the 2D posterior distributions from Planck on the $6$ cosmological parameters assuming either VM either a cosmological constant as dark energy component. The difference in the parameters is mostly associated with the geometric degeneracy in the distance to CMB last scattering. 

We also report in Table~\ref{tablefix} the constraints for the VM scenario from the combined Planck+R16 dataset. However, as we can notice from the last line of the table, where we report the mean minus log likelihoods, ${\bar{\chi}}^2_{\rm eff}$, the inclusion of the R16 prior, that consists in one single data point, results in an increase of $\Delta {\bar{\chi}}^2_{\rm eff} \sim 8$, clearly showing a tension between the Planck data and the R16 prior also in the VM scenario. It is however worth noticing that while the Planck dataset alone in the case of a cosmological constant gives ${\bar{\chi}}^2_{\rm eff}=12967.69$ (see \cite{planck2015}) here we get ${\bar{\chi}}^2_{\rm eff}=12964.64$ for VM, providing a better fit to the same dataset with $\Delta {\bar{\chi}}^2_{\rm eff} \sim -3$.

When we include in the parameter space the sum of neutrino masses (which must exist), the running of the scalar spectral index, and $\neff$ then VM provides a more consistent picture. The constraints 
on this $9$ parameters VM scenario are reported in Table~\ref{tablefix2} for $3$ data combinations (Planck TT, Planck, Planck+R16) and also, for comparison, for the cosmological constant scenario for the Planck+R16 case. 

As we can see, in this case we have that the Planck data alone provide the constraint $H_0=76.5^{+2.3}_{-1.9}$ at $68 \%$ C.L., now in agreement with R16. Moreover, the VM model provides a better mean fit over $\lcdm$ by $\Delta{\bar{\chi}}^2_{\rm eff}=-7.57$ and a value of $H_0=74.8\pm1.4$ at $68 \%$ C.L. for the Planck+R16 case.
The shift in $H_0$ also leads to a lowering of the present dimensionless matter density $\om=0.252^{+0.011}_{-0.014}$. The long period of matter domination before the vacuum phase transition enhances growth, and the strongly negative dark energy equation of state means that dark energy density only becomes appreciable at relatively late times. These combine to raise the 
mass fluctuation amplitude to $\sigma_8=0.877^{+0.039}_{-0.031}$ (see Table~\ref{tablefix2}). However, note that the 
weak lensing parameter $S_8=\sigma_8(\om/0.3)^{0.5}$ actually decreases relative to the $\lcdm$ case, from $0.852\pm0.018$ to $0.803\pm0.022$, putting it in better agreement with  
weak lensing results from the Kilo Degree Survey \cite{kids} and Dark Energy Survey \cite{deswl1,deswl2}. Also, the reduced high redshift $H(z)$ may ameliorate tension in the Lyman alpha-quasar cross-correlations (see \cite{1708.02225}). 

As we can see from Table~\ref{tablefix2} the agreement with the R16 prior comes at the expense of a smaller value of the neutrino effective number $\neff$ with respect to the standard $\neff=3.046$ at 
the level of $\sim 1.5 \sigma$. Also the bounds on neutrino masses are weaker with respect to the cosmological constant case, and some hints are present for a neutrino mass such that $\Sigma m_{\nu} \sim 0.27\,$eV, and for a negative running at the level slightly above $1 \sigma$. 
This should be compared with the same $9$ parameters fit under $\Lambda CDM$ reported in the fourth column of Table~\ref{tablefix2} in the case of the Planck+R16 dataset. As we can see, the agreement in this case is obtained  at the expenses of an higher value for $\neff$ at about $1.5 \sigma$, $\neff=3.31\pm0.18$, and with a strong upper limit on the neutrino mass $\Sigma m_{\nu} < 0.07\,$eV at $68 \%$ C.L..

We can therefore state that in the case of a $9$ parameters analysis {\it both\/} a cosmological constant and VM show some needs for extra physics in order to make the Planck data compatible with the R16 prior. This extra physics is mainly connected with the neutrino effective number $\neff$ that should be {\it larger\/} than the expected value when a cosmological constant is assumed and {\it smaller\/} in the case of VM.

However, as also pointed out in the introduction, the Planck data provides a $\sim 2.5 \sigma$ indication for a larger weak lensing CMB spectrum amplitude $\alens$ (see e.g.\ \cite{plancknewtau}). While the nature of this anomaly is still unclear, it is clearly interesting to provide constraints also in a further extended scenario, varying also $\alens$. We report the results of this analysis in Table~\ref{tablefix3}. In this $10$ parameters framework the VM model prefers now a neutrino mass with $\Sigma m_{\nu}=0.51\pm0.23\,$eV at $68 \%$ C.L. while the neutrino effective number is perfectly compatible with the standard value $\neff=3.046$. In the same $10$ parameters framework and for the same Planck+R16 dataset, but assuming a cosmological constant, we found (see the fourth column in Table~\ref{tablefix3}) that there is no preference for a neutrino mass, with a $68 \%$ C.L. upper limit of $\Sigma m_{\nu} < 0.149\,$eV, while we have an indication for $N_{\rm eff}=3.41\pm0.20$ at $68 \%$ C.L., i.e. almost $2 \sigma$ above the standard value. It is therefore clear that in the $10$ parameters framework the VM model offers an important advantage over the cosmological constant since it solves the tension on the Hubble constant without the need of a non-standard value for $N_{\rm eff}$. In practice, the Planck data under a VM model prefers a value of the Hubble constant {\it larger\/} than the R16 value, but this can be alleviated by introducing a neutrino mass that is well in agreement with current laboratory data (see e.g. \cite{capozzi}). It is also worth noticing that the $\alens$ tension seems somewhat alleviated in the VM scenario and that the value of $S_8$ is now in even better agreement with the recent cosmic shear results from the Kilo Degree Survey \cite{kids}.

We however remark that there can be difficulties with other observational data sets not considered here such as redshift space distortions and supernova distances. We leave that for future work. Still, the improvement in $\chi^2$, the defusing of the $H_0$ tension (and possible amelioration of the weak lensing tension), and of course the strong theoretical foundation of the model together with it having no cosmological constant to explain, makes it worthy of further investigation.

%%%%%%%%%%%%%%%%%%%%%%%%%%%%%%%%%%%%%%%%%%%%%% 
\subsection{Elaborated VM (varying $M$)}

%%%%%%%%%%%%%%%%%%%%%%%%%%%%%
\begin{figure*}[htbp!]
\includegraphics[width=\columnwidth]{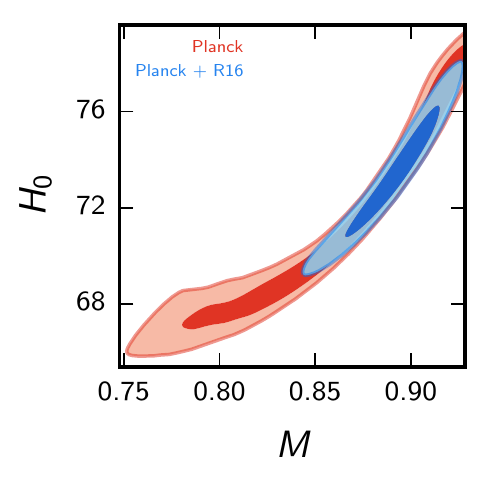} 
\caption{Constraints on the $M$-$H_0$ space of the elaborated VM model from the Planck and Planck+R16 datasets in the $6+1$ parameters analysis.}
\label{M_vs_H0}
\end{figure*}
%%%%%%%%%%%%%%%%%%%%%%%%%

We now consider the more ad hoc VM model that includes a cosmological constant, i.e.\ allow the vacuum criticality parameter $M$ to float.
Constraints are given in Table~\ref{table1} considering a scenario based on $6+1$ cosmological parameters. 
We can immediately see from the Table that allowing $M$ to float lowers the value of the Hubble constant from the Planck data, making it more compatible with the R16 prior, with $H_0=71.5^{+2.8}_{-5.1}$ km/s/Mpc at $68 \%$ C.L.. Considering the Planck+R16 dataset we get $H_0=73.4\pm1.8$ km/s/Mpc at $68 \%$ C.L., with $\Delta{\bar{\chi}}^2_{\rm eff}=-5.6$ with respect with the fixed $M$ model reported in Table~\ref{tablefix}, showing that varying $M$ solves the tensions between Planck and R16.
This can also be clearly seen in Figure~\ref{M_vs_H0} where we plot the 2D posteriors in the $M$ vs $H_0$ plane from the Planck and Planck+R16 datasets. Letting $M$ vary allows for lower values of $H_0$ and the R16 prior is now perfectly compatible with the Planck data.
By comparing the 
${\bar{\chi}}^2_{\rm eff}$ values from the Planck+R16 datasets for the
$9$ parameters case in Table~\ref{tablefix2} and $10$ parameters case
in Table~\ref{tablefix3} we see that allowing $M$ to float solves the $H_0$ tension better than the fixed VM or the cosmological constant model, with the inclusion of one  extra parameter (in this 6+1 scenario without a neutrino mass parameter). 

It is however worthwhile to note that the Planck TT and Planck datasets provide only a lower limit to $M$. Since the maximum theoretical value achievable by $M$ in these runs is given by Eq. \ref{eq:vmom}, corresponding to the fixed $M$ case, this means that the Planck data shows no preference for values of $M$ different from those of the original VM model. This can be also seen by the fact that we have a {\it worse\/} ${\bar{\chi}}^2_{\rm eff}$ value when varying $M$ with respect to the fixed case. In practice, the extra parameter space allowed by varying $M$ is not preferred by the Planck data. 

In Table~\ref{table2} and Table~\ref{table3} we report the constraints obtained on cosmological parameters in the case of a varying $M$ model, adding further extra parameters. In Table~\ref{table2} we include in the analysis also the neutrino effective number $\neff$,
the neutrino mass scale $\Sigma m_{\nu}$, and the running of the spectral index $\frac{dn_s}{d\ln k}$. As we can see, there is now no indication for values different from the standard expectations for these parameters. In particular, the neutrino effective number $\neff$ is now more compatible with $3.046$. 
%A variation in $M$ therefore statistically provides a better solution to the $H_0$ tension than using a smaller value of the neutrino effective number as in the fixed $M$ scenario. 
But the $\chi^2$ improvement does not exceed the number (one) of extra parameters added to the original VM model, and the elaborated model suffers from the usual cosmological constant problem. 

In Table~\ref{table3} we report similar constraints but now also letting the $\alens$ parameter to vary, for a total variation of $11$ parameters. As we can see there is now no indication for extra physics or neutrino mass different from zero as was previously the case for the $M$ fixed model.
In practice, there is no need for extra parameters or additional new physics for solving the $H_0$ tension when varying $M$.

%%%%%%%%%%%%%%%%%%%%%%%%%%%%% 
\begin{table*}
\begin{center}\footnotesize
\scalebox{1}{\begin{tabular}{lccccc}
\hline \hline
         & Planck TT& Planck TT & Planck & Planck\\                     
         & (VM)     &       +R16 (VM) & (VM)& +R16 (VM)\\  
\hline
\hspace{1mm}\\

$\Omega_{\textrm{b}}h^2$& $0.02227\,^{+0.00022}_{-0.00025} $  & $0.02212\,\pm 0.00022$& $0.02225\,\pm 0.00015$ & $0.02219\,\pm 0.00016$  \\
\hspace{1mm}\\

$\Omega_{\textrm{c}}h^2$& $0.1195\,\pm 0.0021$   & $0.1212\,\pm 0.0021$& $0.1198\,\pm 0.0014$ & $0.1206\,\pm0.0015$  \\
\hspace{1mm}\\

$\tau$& $0.075\,\pm 0.020$ & $0.068\,\pm 0.019$& $0.075\,\pm0.018$  & $0.070\,\pm 0.017$   \\
\hspace{1mm}\\

$n_s$& $0.9657\,\pm 0.0061$ & $0.9616\,\pm 0.0060$& $0.9642\,\pm 0.0047$ & $0.9623\,\pm 0.0047$   \\
\hspace{1mm}\\

$\log(10^{10}A_S)$& $3.084\,\pm 0.037$  & $3.073\,\pm 0.036$& $3.085\,\pm0.034$ & $3.077\,\pm 0.032$   \\
\hspace{1mm}\\

$H_0$ &      $78.69\,\pm 0.56$   &  $ 78.22\,\pm 0.58$&      $ 78.61\,\pm 0.38$  &  $ 78.39\,\pm 0.39$   \\
\hspace{1mm}\\

$\sigma_8$   & $ 0.930\,\pm 0.018$     &  $ 0.935\,\pm 0.017$  & $ 0.932\,\pm 0.016$&  $ 0.933\,\pm 0.15$  \\
\hspace{1mm}\\

$S_8$   & $ 0.814\,\pm 0.022$     &  $ 0.829\,\pm 0.022$  & $ 0.817\,\pm 0.022$&  $ 0.823\,\pm 0.017$  \\
\hspace{1mm}\\
\hline
\hspace{1mm}\\
${\bar{\chi}}^2_{\rm eff}$ &  $11279.84$  &  $11287.44$ &  $12964.64$ &  
$12972.18$  \\
\hspace{1mm}\\ 

\hline
\hline

\end{tabular}}
\caption{$68 \%$ c.l.\ constraints on cosmological parameters in the VM scenario for different combinations of datasets.} 
\label{tablefix}
\end{center}
\end{table*}

\begin{table*}
\begin{center}\footnotesize
\scalebox{1}{\begin{tabular}{lccccc}
\hline \hline
         & Planck TT & Planck & Planck & Planck \\                     
              &       +R16 (VM)& (VM) & +R16 (VM)& +R16 ($\Lambda$)  \\  
\hline
\hspace{1mm}\\

$\Omega_{\textrm{b}}h^2$&  $0.02197\,\pm 0.00027$   & $0.02211\,\pm 0.00025$& $0.02194\,\pm 0.00020$ & $0.02257\,\pm 0.00020$  \\
\hspace{1mm}\\

$\Omega_{\textrm{c}}h^2$& $0.1146\,^{+0.0038}_{-0.0043}$  & $0.1175\,\pm 0.0034$ & $0.1160\,\pm 0.0031$& $0.1223\,\pm 0.0031$    \\
\hspace{1mm}\\

$\tau$&$0.080\,\pm 0.022$   & $0.078\,\pm 0.020$& $0.076\,\pm 0.019$   & $0.093\,\pm 0.019$\\
\hspace{1mm}\\

$n_s$& $0.941\,\pm 0.012$    & $0.955\,\pm 0.012$& $0.9471\,\pm 0.0093$& $0.9778\,\pm 0.0084$  \\
\hspace{1mm}\\

$\log(10^{10}A_S)$& $3.081\,\pm 0.046$    & $3.085\,\pm 0.041$ & $3.078\,\pm 0.040$& $3.127\,\pm 0.037$  \\
\hspace{1mm}\\

$H_0$ &       $ 74.0\,\pm 1.6$ &  $ 76.5\,^{+2.3}_{-1.9}$ &  $ 74.8\,\pm 1.4$&  $ 69.7\,\pm 1.3$   \\
\hspace{1mm}\\

$\sum m_{\nu}$ [eV] &       $<0.534$  & $<0.503$ & $0.27\,^{+0.10}_{-0.25}$& $<0.14$  \\
\hspace{1mm}\\

$N_{\rm eff}$ &  $2.57\,^{+0.24}_{-0.28}$   &  $2.87\,\pm 0.23$ &  $2.72\,\pm 0.19$ &  $3.31\,\pm 0.18$   \\
\hspace{1mm}\\

$\frac{dn_s}{d\ln k}$ &  $-0.0154\,\pm 0.0099$  &  $-0.0065\,\pm 0.0079$ &  $-0.0091\,\pm0.0078$ &  $-0.0008\,\pm0.0078$  \\
\hspace{1mm}\\

$\sigma_8$  & $ 0.887\,_{-0.029}^{+0.041}$   &  $ 0.900\,_{-0.025}^{+0.043}$ &  $ 0.877\,_{-0.031}^{+0.039}$ 
&  $ 0.850\,\pm 0.019$\\
\hspace{1mm}\\

$S_8$   & $ 0.816\,\pm 0.026$     &  $ 0.809\,\pm 0.020$  & $ 0.803\,\pm 0.022$&  $ 0.849\,\pm 0.018$  \\

\hspace{1mm}\\
\hline
\hspace{1mm}\\
${\bar{\chi}}^2_{\rm eff}$ &  $11281.84$  &  $12967.07$ &  $12968.69$ &  
$12976.26$  \\
\hspace{1mm}\\
\hline
\hline

\end{tabular}}
\caption{$68 \%$ c.l.\ constraints on cosmological parameters in the VM scenario, including $\sum m_{\nu}$ + $N_{\rm eff}$ + $\frac{dn_s}{d\ln k}$, for different combinations of datasets. For comparison, in the fourth, last, column we report the constraints assuming a cosmological constant for the Planck+R16 dataset. If only upper limits are shown, they are at 95\% C.L..} 
\label{tablefix2}
\end{center}
\end{table*}

\begin{table*}
\begin{center}\footnotesize
\scalebox{1}{\begin{tabular}{lccccc}
\hline \hline
        & Planck TT& Planck & Planck & Planck\\                     
              &   +R16 (VM) &(VM) & +R16 (VM)& +R16 ($\Lambda$) \\  
\hline
\hspace{1mm}\\

$\Omega_{\textrm{b}}h^2$&  $0.02228\,\pm 0.00031$    & $0.02231\,\pm 0.00028$& $0.02214\,\pm 0.00022$& $0.02278\,\pm 0.00022$\\
\hspace{1mm}\\

$\Omega_{\textrm{c}}h^2$& $0.1158\,^{+0.0042}_{-0.0047}$   & $0.1187\,\pm 0.0036$ & $0.1172\,\pm0.0032$  & $0.1222\,\pm0.0031$\\
\hspace{1mm}\\

$\tau$&  $0.064\,^{+0.022}_{-0.025}$   & $0.059\,\pm 0.022$& $0.058\,^{+0.021}_{-0.024}$ & $0.059\,^{+0.021}_{-0.021}$   \\
\hspace{1mm}\\

$n_s$&  $0.959\,\pm 0.016$   & $0.966\,\pm 0.013$& $0.958\,\pm 0.011$
& $0.986\,\pm 0.009$\\
\hspace{1mm}\\

$\log(10^{10}A_S)$& $3.051\,^{+0.045}_{-0.052}$   & $3.050\,\pm 0.044$& $3.043\,^{+0.043}_{-0.049}$ & $3.057\,^{+0.043}_{-0.043}$  \\
\hspace{1mm}\\

$H_0$ & $ 74.6\,\pm 1.6$   &  $ 76.8\,\pm 2.3$ &  $ 74.8\,^{+1.3}_{-1.4}$ &  $ 70.5\,^{+1.4}_{-1.4}$  \\
\hspace{1mm}\\

$\sum m_{\nu}$ [eV] &     $0.54\,^{+0.25}_{-0.35}$ & $<0.829$ & $0.51 \pm 0.23$ & $<0.298$\\
\hspace{1mm}\\

$N_{\rm eff}$ &   $2.85\,^{+0.30}_{-0.37}$  &  $3.04\,\pm 0.26$ &  $2.90\,^{+0.21}_{-0.24}$ &  $3.41\,\pm 0.20$\\
\hspace{1mm}\\

$\frac{dn_s}{d\ln k}$ &  $-0.006\,^{+0.011}_{-0.013}$ &  $0.0001\,\pm 0.0088$ &  $-0.0021\,\pm0.0086$  &  $-0.0049\,\pm0.0078$ \\
\hspace{1mm}\\

$\alens$ &   $1.22\,_{-0.14}^{+0.12}$ &  $1.17\,_{-0.11}^{+0.09}$ &  $1.17\, \pm 0.10$ &  $1.22\,_{-0.097}^{+0.085}$  \\
\hspace{1mm}\\

$\sigma_8$     & $ 0.803\,\pm 0.058$   &  $ 0.841\,_{-0.052}^{+0.064}$ &  $ 0.811\,_{-0.055}^{+0.047}$ &  $ 0.806\,_{-0.033}^{+0.024}$ \\
\hspace{1mm}\\

$S_8$   & $ 0.745\,\pm 0.046$     &  $ 0.761\,\pm 0.037$  & $ 0.752\,\pm 0.035$&  $ 0.798\,\pm 0.026$  \\
\hspace{1mm}\\
\hline
\hspace{1mm}\\
${\bar{\chi}}^2_{\rm eff}$ &  $11280.36$  &  $12965.40$ &  $12966.29$ &  
$12971.19$  \\
\hspace{1mm}\\

\hline
\hline

\end{tabular}}
\caption{$68 \%$ c.l.\ constraints on cosmological parameters in 
the VM scenario, including $\sum m_{\nu}$ + $N_{\rm eff}$ + $\frac{dn_s}{d\ln k}$ + $\alens$, 
for different combinations of datasets. 
For comparison, on the fourth, last, column we report the constraints assuming a cosmological constant for the Planck+R16 dataset. If only upper limits are shown, they are at 95\% c.l..} 
\label{tablefix3}
\end{center}
\end{table*}

%%%%%%%%%%%%%%%%%%%%%%%%%%%%% 
\begin{table*}
\begin{center}\footnotesize
\scalebox{1}{\begin{tabular}{lccccc}
\hline \hline
         & Planck TT& Planck TT& Planck & Planck \\                     
         &     &        +R16  & & +R16  \\  
\hline
\hspace{1mm}\\

$\Omega_{\textrm{b}}h^2$& $0.02224\,\pm 0.00024 $& $0.02224\,\pm 0.00023$    & $0.02225\,\pm 0.00016$& $0.02224\,\pm 0.00016$  \\
\hspace{1mm}\\

$\Omega_{\textrm{c}}h^2$& $0.1197\,\pm 0.0022$& $0.1197\,\pm 0.0022$    & $0.1199\,\pm 0.0014$ & $0.1199\,\pm0.0014$    \\
\hspace{1mm}\\

$\tau$& $0.078\,\pm 0.019$& $0.077\,\pm0.020$   & $0.078\,\pm 0.017$& $0.078\,\pm 0.017$    \\
\hspace{1mm}\\

$n_s$& $0.9656\,\pm 0.0062$& $0.9657\,\pm 0.0062$    & $0.9644\,\pm 0.0048$& $0.9643\,\pm 0.0047$    \\
\hspace{1mm}\\

$\log(10^{10}A_S)$& $3.089\,\pm 0.037$& $3.087\,\pm0.037$    & $3.092\,\pm 0.033$& $3.090\,\pm 0.033$    \\
\hspace{1mm}\\

$H_0$ &      $71.5\,^{+2.8}_{-5.1}$&      $ 73.3\,\pm 1.9$    &  $ 71.6\,^{+2.8}_{-5.1}$ &  $ 73.4\,\pm 1.8$   \\
\hspace{1mm}\\

$M$ &      $>0.785$&      $0.889\,_{-0.012}^{+0.022}$   & $>0.789$ & $0.891\,_{-0.012}^{+0.019}$  \\
\hspace{1mm}\\

$\sigma_8$   & $ 0.867\,_{-0.048}^{+0.028}$   & $ 0.883\,\pm 0.025$   &  $ 0.870\,_{-0.045}^{+0.028}$ &  $ 0.886\,\pm 0.021$  \\
\hspace{1mm}\\

$S_8$   & $ 0.836\,\pm 0.026$     &  $ 0.831\,\pm 0.022$  & $ 0.838\,\pm 0.022$&  $ 0.833\,\pm 0.017$  \\
\hspace{1mm}\\

\hline

\hspace{1mm}\\
${\bar{\chi}}^2_{\rm eff}$ &  $11281.18$  &  $11281.98$ &  $12966.06$ &  
$12966.59$  \\
\hspace{1mm}\\

\hline
\hline

\end{tabular}}
\caption{$68 \%$ c.l.\ constraints on cosmological parameters in the elaborated VM scenario, for different combinations of datasets.
If only lower limits are shown, they are at 95\% C.L..} 
\label{table1}
\end{center}
\end{table*}

\begin{table*}
\begin{center}\footnotesize
\scalebox{1}{\begin{tabular}{lccccc}
\hline \hline
         & Planck TT& Planck TT& Planck & Planck \\                     
         &      &      +R16&  & +R16  \\  
\hline
\hspace{1mm}\\

$\Omega_{\textrm{b}}h^2$& $0.02195\,^{+0.00040}_{-0.00046} $& $0.02226\,^{+0.00032}_{-0.00038}$   & $0.02206\,\pm 0.00025$& $0.02212\,^{+0.00022}_{-0.00025}$   \\
\hspace{1mm}\\

$\Omega_{\textrm{c}}h^2$& $0.1155\,\pm 0.0054$  & $0.1183\,^{+0.0046}_{-0.0053}$ & $0.1175\,\pm0.0033$  & $0.1180\,\pm 0.0033$  \\
\hspace{1mm}\\

$\tau$& $0.083\,\pm 0.023$   & $0.086\,\pm 0.022$& $0.082\,\pm0.019$& $0.080\,\pm 0.019$   \\
\hspace{1mm}\\

$n_s$& $0.940\,\pm 0.024$   & $0.959\,^{+0.017}_{-0.021}$& $0.953\,\pm 0.011$ & $0.957\,\pm 0.011$  \\
\hspace{1mm}\\

$\log(10^{10}A_S)$& $3.090\,\pm 0.050$   & $3.104\,\pm 0.046$& $3.093\,\pm0.039$  & $3.090\,\pm 0.040$  \\
\hspace{1mm}\\

$H_0$ &      $66.1\,^{+5.2}_{-6.4}$ &  $ 73.0\,\pm 1.7$ &      $ 68.6\,^{+3.9}_{-5.3}$&  $ 73.2\,\pm 1.7$   \\
\hspace{1mm}\\

$M$ &      $>0.742$  & $0.892\,^{+0.030}_{-0.008}$ &      $>0.754$ & $0.899\,^{+0.020}_{-0.009}$   \\
\hspace{1mm}\\

$\sum m_{\nu}$ [eV] &      $<0.640$&      $<0.456$  & $<0.573$ & $<0.428$  \\
\hspace{1mm}\\

$N_{\rm eff}$ &  $2.61\,^{+0.42}_{-0.49}$   &  $2.93\,^{+0.33}_{-0.43}$ &  $2.84\,\pm0.22$ &  $2.90\,\pm 0.21$   \\
\hspace{1mm}\\

$\frac{dn_s}{d\ln k}$ &  $-0.016\,\pm 0.012$ &  $-0.009\,\pm 0.011$  &  $-0.0083\,\pm 0.0081$ &  $-0.0064\,\pm0.0078$   \\
\hspace{1mm}\\

$\sigma_8$   & $ 0.816\,\pm 0.057$      &  $ 0.876\,_{-0.028}^{+0.035}$ & $ 0.830\,_{-0.047}^{+0.054}$&  $ 0.875\,_{-0.030}^{+0.024}$ \\
\hspace{1mm}\\

$S_8$   & $ 0.845\,\pm 0.031$     &  $ 0.827\,\pm 0.024$  & $ 0.836\,\pm 0.026$&  $ 0.822\,\pm 0.022$  \\
\hspace{1mm}\\

\hline

\hline

\hspace{1mm}\\
${\bar{\chi}}^2_{\rm eff}$ &  $11282.83$  &  $11282.81$ &  $12968.56$ &  
$12968.02$  \\
\hspace{1mm}\\

\hline
\hline

\end{tabular}}
\caption{$68 \%$ c.l.\ constraints on cosmological parameters in 
the elaborated VM scenario, including $\sum m_{\nu}$ + $N_{\rm eff}$ + $\frac{dn_s}{d\ln k}$, 
for different combinations of datasets. If only lower limits are shown, they are at 95\% c.l..} 
\label{table2}
\end{center}
\end{table*}

\begin{table*}
\begin{center}\footnotesize
\scalebox{1}{\begin{tabular}{lccccc}
\hline \hline
         & Planck TT& Planck TT& Planck & Planck \\                     
         &      &     +R16 & & +R16  \\  
\hline
\hspace{1mm}\\

$\Omega_{\textrm{b}}h^2$& $0.02287\,^{+0.00056}_{-0.00069} $& $0.02297\,^{+0.00045}_{-0.00039}$    & $0.02227\,\pm 0.00028$& $0.02239\,\pm 0.00027$  \\
\hspace{1mm}\\

$\Omega_{\textrm{c}}h^2$& $0.1214\,^{+0.0056}_{-0.0071}$& $0.1218\,^{+0.0050}_{-0.0058}$   & $0.1188\,\pm 0.0035$ & $0.1195\,\pm0.0034$  \\
\hspace{1mm}\\

$\tau$& $0.066\,^{+0.023}_{-0.026}$& $0.067\,^{+0.022}_{-0.026}$   & $0.059\,\pm 0.021$& $0.060\,\pm 0.021$    \\
\hspace{1mm}\\

$n_s$& $0.991\,^{+0.030}_{-0.035}$& $0.996\,^{+0.023}_{-0.019}$   & $0.964\,\pm 0.013$& $0.969\,\pm 0.012$  \\
\hspace{1mm}\\

$\log(10^{10}A_S)$& $3.067\,^{+0.048}_{-0.055}$& $3.070\,^{+0.047}_{-0.053}$   & $3.048\,\pm 0.044$& $3.052\,\pm 0.043$   \\
\hspace{1mm}\\

$H_0$ &      $71.6\,^{+6.6}_{-7.7}$&      $ 73.2\,\pm 1.7$   &  $ 67.2\,^{+3.8}_{-5.4}$ &  $ 72.9\,\pm 1.7$   \\

\hspace{1mm}\\

$M$ &      $>0.754$&      $0.856\,^{+0.050}_{-0.031}$  & $>0.724$ & $0.889\,^{+0.027}_{-0.010}$  \\
\hspace{1mm}\\

$\sum m_{\nu}$ [eV] &      $0.54\,^{+0.18}_{-0.50}$&      $<1.14$ & $0.51\,^{+0.20}_{-0.44}$ & $<0.847$ \\
\hspace{1mm}\\

$N_{\rm eff}$ &  $3.50\,^{+0.54}_{-0.75}$ &  $3.57\,^{+0.43}_{-0.50}$  &  $3.04\,\pm 0.25$ &  $3.11\,^{+0.24}_{-0.26}$ \\
\hspace{1mm}\\

$\frac{dn_s}{d\ln k}$ &  $0.005\,\pm 0.015$ &  $0.007\,^{+0.012}_{-0.014}$ &  $-0.0005\,\pm 0.0089$ &  $0.0010\,\pm0.0082$   \\
\hspace{1mm}\\

$\alens$ &  $1.35\,_{-0.17}^{+0.14}$ &  $1.35\,_{-0.17}^{+0.13}$ &  $1.22\,_{-0.11}^{+0.10}$ &  $1.195\, \pm 0.096$   \\

\hspace{1mm}\\

$\sigma_8$   & $ 0.748\,_{-0.072}^{+0.080}$   & $ 0.764\,_{-0.054}^{+0.070}$   &  $ 0.745\,\pm 0.068$ &  $ 0.807\,_{-0.044}^{+0.051}$  \\
\hspace{1mm}\\

$S_8$   & $ 0.739\,\pm 0.051$     &  $ 0.736\,\pm 0.047$  & $ 0.772\,\pm 0.037$&  $ 0.769\,\pm 0.037$  \\

\hspace{1mm}\\
\hline

\hspace{1mm}\\
${\bar{\chi}}^2_{\rm eff}$ &  $11279.46$  &  $11279.66$ &  $12965.91$ &  
$12966.30$  \\
\hspace{1mm}\\

\hline
\hline

\end{tabular}}
\caption{$68 \%$ c.l.\ constraints on cosmological parameters in 
the elaborated VM scenario, including $\sum m_{\nu}$ + $N_{\rm eff}$ + $\frac{dn_s}{d\ln k}$ + $\alens$, 
for different combinations of datasets. If only upper or lower limits are shown, they are at 95\% c.l..} 
\label{table3}
\end{center}
\end{table*}

%%%% 
A summary comparing the $\chi^2$ of the VM models 
with $\lcdm$ is given in Table~\ref{chi2}.

%%%%%%%%%%%%%%%%%%%%%%%%% 
\begin{table} 
\begin{center} 
\begin{tabular}{l|c|c} 
Model & $\Delta N_{\rm par}$ & $\Delta{\bar{\chi}}^2_{\rm eff}$ \\ 
\hline  
{\bf Planck only, minimal $6$ parameters:} & & \\ 
$\Lambda$CDM & $-$ & $-$ \\ 
Vacuum Metamorphosis\ \ & 0 & $-3.05$ \\ 
VM elaborated & 1 & $-1.63$ \\ 
$\alens(\ell)$ & $1^{\star}$ & $-0.39^\star$ \\ 
\hline 
{\bf Planck only, +$m_\nu$,$\neff$,$\frac{dn_s}{d\ln k}$:} & & \\ 
$\Lambda$CDM & $-$ & $-$ \\ 
Vacuum Metamorphosis\ \ & 0 & $-1.62$ \\ 
VM elaborated & 1 & $-0.27$ \\ 
\hline 
{\bf Planck+R16, +$m_\nu$,$\neff$,$\frac{dn_s}{d\ln k}$:} & & \\ 
$\Lambda$CDM & $-$ & $-$ \\ 
Vacuum Metamorphosis\ \ & 0 & $-7.57$ \\ 
VM elaborated & 1 & $-8.24$ \\ 
\end{tabular}
\end{center}
\caption{Comparison of the beyond standard physics models with 
standard $\Lambda$CDM. The number of additional parameters 
relative to $\lcdm$ is $\Delta N_{\rm par}$, and the improvement in 
the fit relative to $\lcdm$ is $\Delta{\bar{\chi}}^2_{\rm eff}$. 
The star superscript  
indicates that this model is compared relative to $\lcdm+\alens$, 
for the baseline CMB data plus CMB lensing. 
} 
\label{chi2} 
\end{table}

%%%%%%%%%%%%%%%%%%%%%%%%%%%%% 
\section{Scale Dependent Lensing Amplitude} \label{sec:scale} 

In a second approach to beyond standard physics, we test the ``Planck'' dataset with a scale dependent scaling of the gravitational lensing amplitude. This seeks to explore indications that cosmological parameters derived from the lower multipole ($\ell\lesssim1000$) data and the higher multipole ($\ell\gtrsim1000$) data can differ by $\sim1\sigma$. 
In this case, in addition to the six parameters of the standard $\Lambda$CDM model (VM is not used in this section), we re-parametrize $\alens$ from a constant 
(seventh parameter) to both an amplitude and a slope, giving eight parameters in total. 

Specifically, 
\begin{equation} \label{scaling}
\alens=\alensz\times \left[1+B\,\log_{10}\left(\frac{\ell}{300}\right)\right] \ . 
\end{equation}
This form is motivated by the behavior of various beyond standard scale 
dependent physics, such as modified gravity, neutrino mass, and cold dark energy, investigated in \cite{1507.08292}. The amplitude $\alensz$ is the value at $\ell=300$, in the vicinity of the first acoustic peak, and roughly represents the mean over the full multipole range. The slope $B$ can be positive or negative, and its prior in Table~\ref{priors} prevents $\alens$ from going negative anywhere in the 
multipole range. 

The constraints on $\alensz$ and $B$ are reported in Table~\ref{tablealens} for several combinations of datasets.
The Planck TT and Planck datasets both favor a value for $\alensz$ larger than the expected value, while the $B$ parameter is unconstrained. Comparing to the standard $\lcdm$ case, the parameter values  do not shift appreciably and the $\chi^2$ improves by less than 0.4 (at the cost of 1 more parameter). However we found that these mild shifts are in the right direction to alleviate the several tensions. We found that for the Planck dataset the Hubble constant is now constrained to be $H_0=67.86\pm0.74$ km/s/Mpc at $68 \%$ C.L., i.e. bringing the tension with the R16 prior from $3.24$ standard deviations to
$2.87$. Also the $S_8$ parameter is smaller and now constrained from the Planck dataset to be $S_8=0.818\pm0.024$ at $68 \%$ C.L., in better agreement with cosmic shear measurements.

\begin{figure}
\centering
\includegraphics[width=0.95\columnwidth]{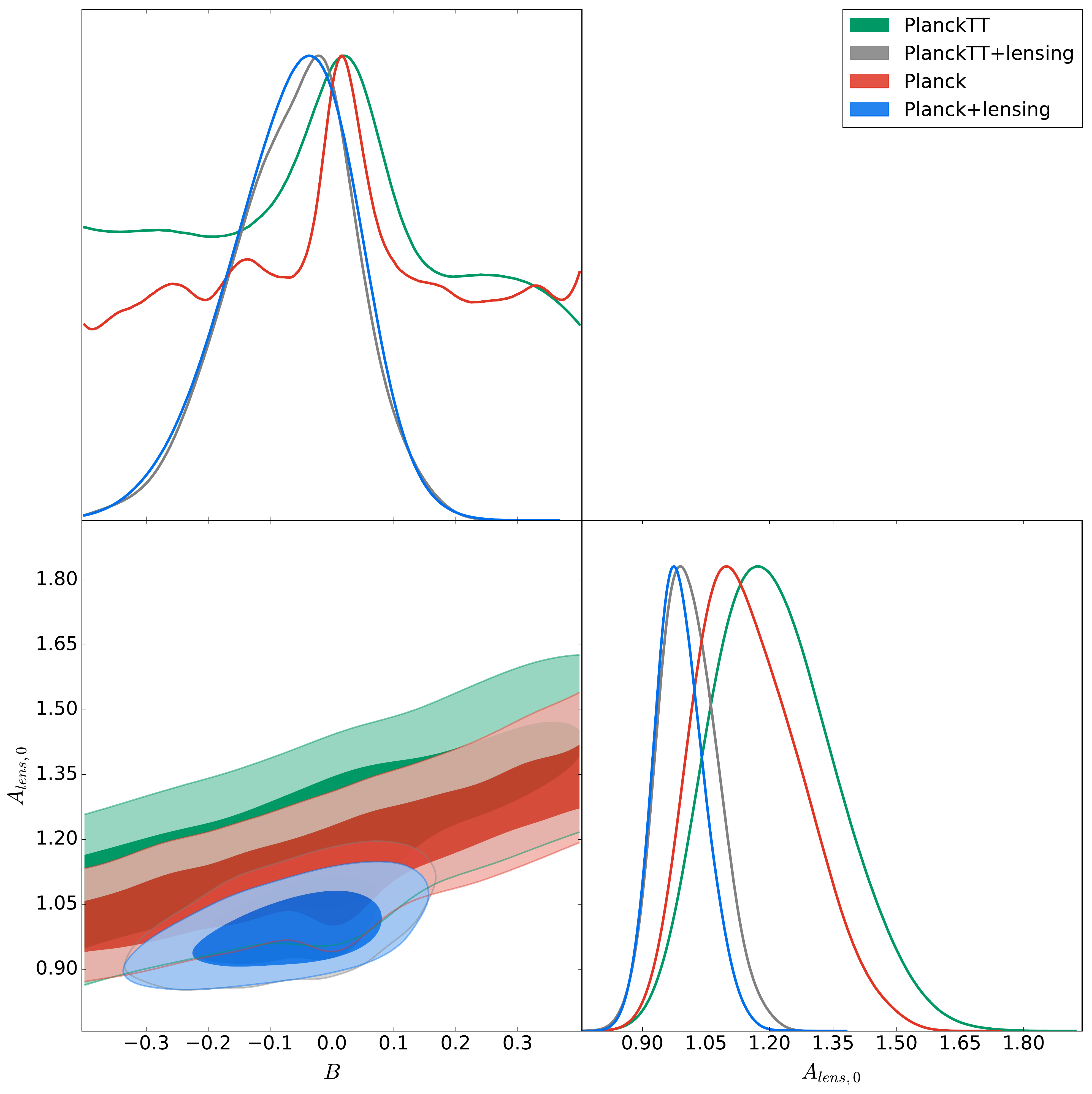} 
\caption{Triangular plot showing the posteriors of $A_{\rm lens,0}$ and $B$ for the datasets considered, as well as their 2D joint confidence contour at 68\% CL.}
\label{figalensB}
\end{figure}

The one additional parameter $B$ 
cannot be determined with the ``Planck'' data set alone. To constrain the scale dependence of the lensing amplitude, we must include CMB lensing data, i.e.\ use the lensing potential power spectrum derived from the CMB 
trispectrum analysis; we refer to this as ``Planck+lensing''. 
Table~\ref{tablealens} summarizes the results, and 
Figure~\ref{figalensB} shows the 1D and 
joint probability distributions of the lensing amplitude parameters. 

The positive correlation between $\alensz$ and $B$ can be understood as preserving the CMB lensing power spectrum amplitude where it has the most power, at $\ell<300$.

%%%%%%%%%%%%%%%%%%%%%%%%%%%%%%%
\begin{table*}
\begin{center}\footnotesize
\scalebox{1}{\begin{tabular}{lccccc}
\hline \hline
         & Planck TT & Planck TT & Planck & Planck \\                     
         &      & +lensing    & & +lensing \\  
\hline
\hspace{1mm}\\

$B$ & \ \ unconstrained\ \ &  $-0.07\,\pm 0.10$ & \ \ unconstrained\ \ &  $-0.076\,^{+0.11}_{-0.099}$  \\
\hspace{1mm}\\

$A_{\rm lens,0}$ &  $1.22\,_{-0.17}^{+0.13}$ &  $1.014\,_{-0.080}^{+0.068}$  & $1.17\,^{+0.12}_{-0.15}$& $0.994\, ^{+0.061}_{-0.061}$  \\
\hspace{1mm}\\
\hline
\hspace{1mm}\\
${\bar{\chi}}^2_{\rm eff}$ & $11276.91$ & $11293.67$ & $12963.55$ & $12979.64$ \\
\hspace{1mm}\\
\hline
\hline

\end{tabular}}
\caption{$68 \%$ c.l.\ constraints on the amplitude and slope of the scale dependent scaling of the gravitational lensing amplitude (Eq.~\ref{scaling}), using different datasets. } 
\label{tablealens}
\end{center}
\end{table*}

%%%%%%%%%%%%%%%%%%%%%%%%%%%%%%

The inclusion of the lensing data brings the value of $\alensz$
back in agreement with the standard value, and it now constrains the slope to $B=-0.076^{+0.11}_{-0.099}$. We find negligible shift in the cosmological  parameters. Thus, this form of scale dependence (linear in $\log\ell$) cannot solve the $H_0$ tension\footnote{Note 
that the scale dependent physics considered in \cite{1507.08292} does lead to a negative value of 
$B\approx-0.015$ for the massive neutrino and cold dark energy cases (while $B$ has a positive sign for the $f(R)$ gravity case). Current experimental precision is insufficient to constrain such scale dependent physics.}.

We remark however, by looking at the last line in Table~\ref{tablealens}, that the inclusion of CMB lensing to the Planck dataset significantly increases the ${\bar{\chi}}^2_{\rm eff}$ by $\sim 16$. Since the CMB lensing consists of about $8$ datapoints, this clearly shows a significant tension between the Planck and lensing datasets that not even a scale dependence for $\alens$ seems able to solve.

%%%%%%%%%%%%%%%%%%%%%%%%%%%
\section{Conclusions} \label{sec:concl}

Current CMB and local Hubble constant data, taken at face value and 
interpreted within a $\lcdm$ cosmological model, show a 
tension in the value of $H_0$. This tension can be removed by taking 
the dark energy not to be near cosmological constant behavior but with 
a very unusual nature -- deeply phantom and rapidly evolving. Rather 
than treating this phenomenologically, we resuscitate the vacuum 
metamorphosis theory of Parker and collaborators, involving a phase 
transition in the nature of gravity and the vacuum, based on calculations 
within quantum gravity. 

We demonstrate that vacuum metamorphosis provides a solution to the $H_0$ tension, and indeed 
yields an improvement in $\chi^2$ by 7.5 over $\lcdm$ with the same number of 
parameters. Moreover, it can also ameliorate possible tension in the weak 
lensing amplitude $S_8$ seen between Planck and some ground based surveys. 
Given the theory's robust foundation and reasonable motivation, including no 
explicit or implicit cosmological constant, it is 
worthwhile to investigate it further in future work, in particular examining 
consistency with further data sets such as baryon acoustic oscillations and 
supernova distances. Note that  analyses (such as the recent 
\cite{Heavens:2017hkr}) based on Bayesian Evidence and that disfavor extensions to the $\Lambda$CDM model on the basis of its ``simplicity'' may obtain different conclusions given the VM model that has the same number of parameters as $\Lambda$CDM. 

Another extension of the standard model involves scale dependence of the 
CMB lensing amplitude $\alens$, beyond what exists in the standard model. 
This has a more modest motivation, from the lesser apparent tension between 
cosmological parameters derived from CMB data at high and low multipoles 
(roughly less than and greater than $\ell\approx1000$). Such scale dependence 
could arise from beyond standard model physics such as modified gravity, 
cold dark energy, or massive neutrinos. We do not find any evidence for a 
tilt in the CMB lensing amplitude, though the Planck lensing data is not 
precise enough to constrain this tightly. 
%These particular physics cases 
%predict a tilt of order $-0.015$ but the uncertainty from Planck data is 
%$\sim7$ times larger. 

Future CMB data from Stage 3 experiments, and particularly from a CMB 
Stage 4 experiment, can continue to test the nature of dark energy, 
beyond standard physics, and consistency between the high and low redshift 
universe. Any solution must fit the rich array of data.  %including from 
%cosmic surveys of large scale structure and supernovae. 
All together will 
evaluate tensions and anomalies and shed light on whether we are seeing 
systematics, statistical excursions, or indeed new physics, perhaps even 
definite signs of quantum gravity. 

\acknowledgments 
 
We thank Robert Caldwell for helpful discussion. 
EDV acknowledges support from the European Research Council in the form 
of a Consolidator Grant with number 681431. 
EL was supported in part by the Energetic Cosmos Laboratory and by 
the U.S.\ Department of Energy, Office of Science, Office of High Energy 
Physics, under Award DE-SC-0007867 and contract no.\ DE-AC02-05CH11231.
AM thanks the University of Manchester and the Jodrell Bank Center for Astrophysics for hospitality.

\end{document}